\documentclass[12pt]{article}
\usepackage{graphicx}
\usepackage{amssymb}

\title{Unifying Local  Dynamics\\ in\\ Two-State Spin Systems}
\author{Serge Galam \\
Universit\'e Pierre et Marie Curie et CNRS,\\
Laboratoire des Milieux D\'esordonn\'es et H\'et\'erog\`enes\\
Case 86, 4 place Jussieu, F-75252 Paris Cedex 05, France }
\begin{document}
\date{(galam@ccr.jussieu.fr)}
\maketitle

\begin{abstract}
We present a new two-state $\{\pm\}$ opinion dynamics model which defines a general frame to include all local  dynamics in two-state spin systems. Agents evolve by probabilistic local rules. In each update, groups of various sizes $k$ are formed according to some probability distribution $\{a_k\}$. Given a specific group with an initial $(j)$ agents sharing opinion $\{+\}$ and $(k-j)$ agents opinion $\{-\}$,  all $k$ members adopt  opinion $\{+\}$ with a probability $m_{k,j}$ and opinion $\{-\}$ with $(1-m_{k,j})$. A very rich and new spectrum of dynamics is obtained.  The final opinion is a polarization along the initial majority, along  the initial minority or a perfect consensus with an equality of opinions as function of the parameters $\{a_k, m_{k,j}\}$. In last case, two regimes exist, monotonic and dampened oscillatory. The transition from polarization to consensus dynamics occurs for values of the parameters which reproduce exactly the Voter model. A scheme is presented to express any local update in terms of a specific set $\{a_k, m_{k,j}\}$. Most existing opinion models are exhibited at particular limits.

\end{abstract}



\newpage

In this Letter we present a new two-state opinion dynamics model which creates a general frame to unify all dynamics using local updates. It exhibits a rather rich and complex dynamical behavior. Any local update is shown to correspond to a specific set of parameters. Most  of existing opinion dynamics models are thus obtained at particular limits. The model consists of a group of agents sharing either one of two opinions represented by two-state $\{\pm\}$ opinion variables like Ising spins.
In each update, agents are randomly distributed among groups of various sizes $k$ according to some probability distribution $\{a_k\}$ with the constraint,
\begin{equation}
\sum_{k=1}^L a_k=1 ,
\end{equation}
where $k=1, 2, ..., L$  stands for groups
of respective sizes $1, ..., L$ with $L$ being the largest size. Then within each  group of size $k$ with $j$ agents sharing opinion $\{+\}$ and $(k-j)$ agents opinion $\{-\}$,  all $k$ members adopt  opinion $\{+\}$ with a probability $m_{k,j}$ and opinion $\{-\}$ with $(1-m_{k,j})$. Accordingly, after one update, the probability $p(t+1)$ to find at random an agent sharing opinion $\{+\}$ writes,
\begin{equation}
p(t+1)=\sum_{k=1}^L a_k \sum_{j=0}^k m_{k,j} C_k^j p(t)^{j} [1-p(t)]^{k-j}          \ ,
\end{equation}
where $C_k^j\equiv  \frac{k!}{j! (k-j)!}$ are binomial coefficients and $p(t)$ is the probability of finding an agent sharing opinion  $\{+\}$ picked up randomly before the update.

Our model represents a natural outgrowth of  recent work on Minority opinion spreading in random geometry \cite{mino} combined with another recent approach which suggests that Contrarian behavior produces the scenario of hung elections  \cite{contra}. Minority spreading is obtained taking $m_{k,j}=1$ for $j\geq\frac{k+2}{2}$ and $m_{k,j}=0$ for $j< \frac{k+2}{2}$ in Eq. (2). It corresponds to local deterministic majority rules with all tie cases at even groups yielding an opinion  $\{-\}$. Contrarian behavior   is recovered by considering only odd size groups, i.e. $a_k=0$ for $k$ even, in addition to  $m_{k,j}=1-m$ for $j\geq \frac{k+1}{2}$ and $m_{k,j}=m$ for $j<\frac{k+1}{2}$ where $m$ is the proportion of contrarians. 

Earlier voting models by Galam \cite{voting-all} which considered only one group size $r$ at a time are also recovered from Eq. (2) by taking $a_k=0$ for $k\neq r$,  $a_r=1$, $m_{r,j}=1$ for 
$j\geq \frac{r+2}{2}$ and $m_{r,j}=0$ for $j<\frac{r+2}{2}$. For $r$ odd, it means one  group of arbitrary size with majority rules while for $r$ even, it has in addition  a tie effect biased in favor of the opinion  $\{-\}$. The case of a probabilistic bias $b$ at a tie was also studied in \cite{jan} and is obtained here by taking $m_{r,j}=1$ for 
$j> \frac{r}{2}$, $m_{r,j}=0$ for $j<\frac{r}{2}$ and $m_{r,j}=b$ for $j=\frac{r}{2}$.

Likewise the majority rule model by Krapivsky and Redner \cite{red-1} which is a $r=3$ Galam voting model is readily obtained with $m_{3,3}=m_{3,2}=1$ and $m_{3,1}=m_{3,0}=0$. Its  majority minority extension  by Mobilia and Redner \cite{red-2} is realized with $m_{3,3}=1$, $m_{3,2}=q$, $m_{3,1}=1-q$ and $m_{3,0}=1$.

As seen from above  it is worth noticing that all existing models assume a symmetry between the two opinions, i.e., $m_{k,j}=1-m_{k,k-j}$ except the model of tie driven bias which takes  for even groups $m_{k,\frac{k}{2}}\neq 1-m_{k,\frac{k}{2}}$ \cite{mino}. In contrast Eq. (2) incorporates all possible asymmetries. It represents a unified frame to study a very large number of peculiar dynamics.

This work subscribes to the field of Sociophysics which started more than twenty years ago but stayed limited to a few very physicists while opposed or ignored by everyone else \cite{socio,strike}. Only in the last years did Sociophysics became accepted by the physics community and is attracting a growing number of researchers  \cite{frank-book,weidlich-book,stauffer,neigbhor,huber},

The rest of the Letter is organized as follows. First the dynamics properties of Eq. (2) are studied in details for the  case of symmetric groups of size $4$. A very rich and new spectrum of dynamics is obtained with the occurrence of a phase transition. For some range of the parameters $\{a_4,m_{4,j}\}$, the final opinion is always a polarization along the initial majority. However, depending on what happens at a tie for even groups, the polarization may happen to be along  the initial minority. On the other hand,  for other range of parameters, the dynamics is reversed towards a perfect consensus with an equality of opinions. In this case, two regimes exist, monotonic and dampened oscillatory. The transition from polarization to consensus dynamics is found to occur for values of the parameters which reproduce exactly the Voter model \cite{voter}. 

A scheme is presented to express any local update in terms of a specific set of parameters $\{a_k, m_{k,j}\}$.
On this basis we revisit the various versions of the Sznajd model \cite{sznajd,stauffer} to find out they are peculiar limits of Eq. (2). While its various versions have been  always considered as similar we show that indeed some correspond to the phase with a polarization along the initial majority while another is different in nature since it corresponds to a  Voter models as demonstrated earlier  by Behera and Schweitzer \cite{frank-voter}. 

Considering groups of one unique size $4$ we have  $a_k=0$ for $k\neq 4$ and $a_4=1$ which reduce Eq.(2) to,
\begin{equation}
p(t+1)=    \begin{array}[t]{l}  m_ {4,4}p(t)^4+4 m_ {4,3}p(t)^3  [1-p(t)]+6m_{4,2} p(t)^2  [1-p(t)]^2
 \\ \\ +4m_ {4,1}p(t)  [1-p(t)]^3+m_{4,0} [1-p(t)]^4 \ ,   \end{array}
\end{equation}
which embeds the typical case which exhibits the  tie effect at $m_{4,4}=m_{4,3}=1$ and $m_{4,2}=m_{4,1}=m_{4,0}=0$. It yields a threshold to win public opinion at $77\%$ \cite{voting-all}. 

We start  first with a totally symmetric situation between both opinions, i.e., $m_{4,0}=1-m_{4,4}$, $m_{4,1}=1-m_{4,3}$ and $m_{4,2}=1-m_{4,2}$. From last equality $m_{4,2}=\frac{1}{2}$ which excludes any bias effect at the tie. We also make $m_{4,0}=1-m_{4,4}=0$ which results in,
\begin{equation}
p(t+1)=     \begin{array}[t]{l}  p(t)^4+4m_{4,3}p(t)^3  [1-p(t)]+3 p(t)^2  [1-p(t)]^2
 \\ \\  +4(1-m_{4,3})p(t)  [1-p(t)]^3 \ .  \end{array} 
\end{equation}

Eq. (4) has three fixed point at $0, \frac{1}{2}, 1$ whose respective stability is a function of  $m_{4,3}$. The variation of the flow dynamics is  shown in Fig. (1) for the whole range $0\leq m_{4,3}\leq 1$.
It  yields a phase transition  between a phase with total polarization and a phase with perfect consensus. Various regimes are exhibited in the flow of opinion while reaching an attractor.  It is worth stressing that we are using the words polarization and consensus as in social sciences, i.e., polarization = all agents share the same opinion and  consensus = agents are perfectly distributed between the two opinions \cite{mosco-1}. At odd, physicists used their other way around \cite{sznajd,stauffer,frank-voter}

For $\frac{3}{4}< m_{4,3}\leq 1$, $0$ and $1$ are stable, i.e., attractors of the dynamics, and $\frac{1}{2}$ unstable is the separator. Accordingly any initial proportion of  $\{+\}$ lower (greater) than fifty percent ends up to zero (one) with a polarization of the population along the minus (plus) opinion.  The flow is rapid and monotonic as seen in Fig. (1). 

However for $m_{4,3}<\frac{3}{4}$ the dynamics is reversed with former separator $\frac{1}{2}$ being the unique attractor. Any initial proportion of  $\{+\}$ and  $\{-\}$ ends at an exactly equality of both opinions. However in this phase the local dynamics stays active with agents shifting opinions within a perfectly collective consensus. In contrast in the precedent polarization phase the local dynamics stops when reaching one of the two attractors. Another difference is in the flow regime to reach stability. While it is always monotonic in the polarization phase it is twofold for the consensus phase. It is monotonic for $\frac{1}{4}<m_{4,3}<\frac{3}{4}$ and becomes dampened oscillatory when  $0\leq m_{4,3}<\frac{1}{4}$. As shown in Fig. (1), in that regime in the vicinity of the attractor,  each update shifts the majority from one opinion onto the other. During these successive shifts of majority, the absolute difference between the majority and the minority is reduced at each step before reaching zero at the fifty percent attractor where we have an equal number of $\{+\}$ and $\{-\}$.
These oscillations may be a new instrumental key to understand some two-party system elections  in addition to the earlier contrarian proposal \cite{contra}.

\begin{figure}[h]
\begin{center}
\includegraphics[scale=1]{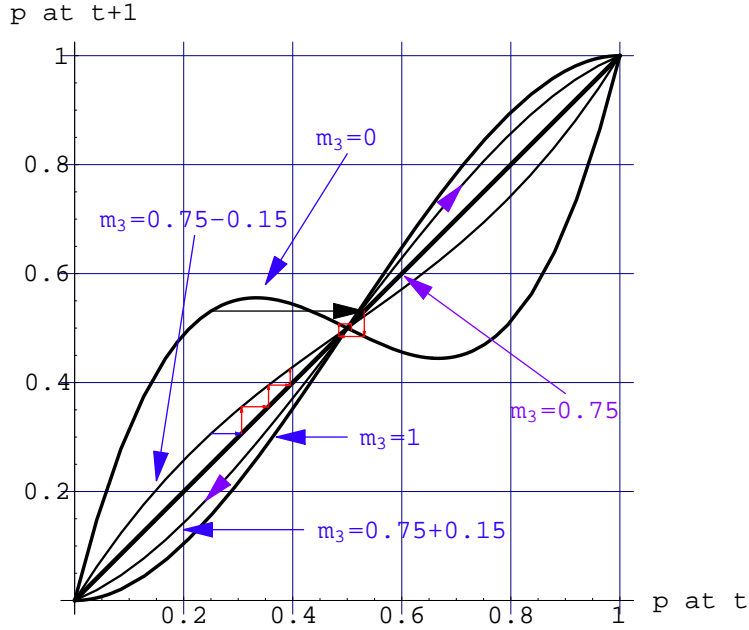}
\caption{Variation of the flow dynamics as function of $m_{4,3}$ in the case of size $4$ ($a_4=1$) with $m_{4,0}=1-m_{4,4}=0$ and $m_{4,2}=\frac{1}{2}$  .  The transition occurs at $m_{4,3}=\frac{3}{4}=0.75$ where the Voter model is recovered. The cases $m_{4,3}=1$ (original Galam) and  the extreme case of a systematic minority convincing power at $m_{4,3}=0$ are shown together with two intermediate cases below and above the critical line at $m_{4,3}=0.75\pm 0.15$. A new feature appears here when $0\leq m_{4,3}<\frac{1}{4}$ with the occurrence of oscillatory shifts between a majority of $\{+\}$ and a majority of $\{-\}$  as shown for the case $m_{4,3}=0$.}
\end{center}
\end{figure}

The transition between the polarization and the consensus phases occurs at $m_{4,3}=\frac{3}{4}$ which drives a qualitative change of the dynamics. Eq. (4) now becomes,
\begin{equation}
p(t+1)=     p(t)^4+3p(t)^3  [1-p(t)]+3 p(t)^2  [1-p(t)]^2+p(t)  [1-p(t)]^3 \ , 
\end{equation}
which corresponds exactly to a Voter model \cite{voter} with $m_{k,j}=\frac{j}{k}$. Here with four neighbors it yields $m_{4,4}=\frac{4}{4}=1$, $m_{4,3}=\frac{3}{4}$, $m_{4,2}=\frac{2}{4}=\frac{1}{2}$, $m_{4,1}=\frac{1}{4}$ and $m_{4,0}=\frac{0}{4}=0$. Moreover expanding Eq. (5) gives $p(t+1)=p(t)$ which expresses the fact that the dynamics keeps unchanged the proportion of  $\{+\}$ and $\{-\}$ Such a conservation of the order parameter is a basic feature of the Voter model. It implies for simulations that the probability that the system eventually ends with all plus spins equals the initial density of plus spins in all spatial dimensions.
All different cases are exhibited in Fig. (1).

 It is worth noticing that a similar phase transition into a consensus phase was also obtained independently by  Mobilia and Redner \cite{red-2} and Galam \cite{contra}. In the first case groups of size $3$ are used under the conditions  $m_{3,3}=1$, $m_{3,2}=q$, $m_{3,1}=1-q$ and $m_{3,0}=1$. There, the transition occurs at
$m_{3,2}=  \frac{2}{3}$ where the recover of the Voter model is noticed. The second case is of a different symmetry since  it uses the conditions  $m_{k,j}=1-m$ for $j\geq \frac{k+1}{2}$ and $m_{k,j}=m$ for $j<\frac{k+1}{2}$ for  odd size  groups, i.e. $a_k=0$ for $k$ even with $m$ the proportions of contrarians. In particular the fixed point $0$ and $1$ are shifted towards $\frac{1}{2}$ as function of $m$ and the Voter model is not recovered at the transition. The oscillatory regime is absent from both cases.

We are now in a position to define a scheme to express any local update in terms of a specific set of the parameters $\{a_k, m_{k,j}\}$. We illustrate it revisiting the Sznajd model \cite{sznajd}. The original Sznajd model at one dimension  considers an Ising spin
model with periodic boundary conditions. Each spin $S_i$ is located at  a lattice site $i = 1, ...,N$ and $S_i=\pm 1$ refer respectively to two opposite opinions.
Then the spins are updated according to the following two step rule. First select two neighboring spins at sites sites $(i)$ and $(i+1)$. 
Second,  if they have the same opinion, i.e. if  $S_i S_{i+1}=1$,
then the two neighbouring sites $(i-1)$ and $(i+2)$ adopt their common opinion. Otherwise, when they have opposite opinions,   i.e. if  $S_i S_{i+1}=-1$, the two neighbouring spins $S_{i-1}$ and $S_{i+2}$ adjust to create an anti-ferromagnetic like ordering.
These rules write,\\

$[a]: (++++),\ (+++-),\ (-+++),\ (-++-)\rightarrow (++++)$,\\

$[b]: (+--+),\ (+---),\ (---+),\ (----)\rightarrow (----)$,\\

$[c]: (++-+),\ (++--),\ (-+-+),\ (-+--)\rightarrow  (-+-+)$,\\

$[d]: (+-++),\ (+-+-),\ (--++),\ (--+-)\rightarrow (+-+-)$,\\ \\
motivated by the social claim ``United we stand, divided we fall''  \cite{socio}. To cast above rules within our frame, we calculate the probability of each configuration and its weight contribution to producing a state $\{+\}$. For instance, $p(t))$ being the proportion of  $\{+\}$,  the configuration $(++-+)$ has a probability of $p(t)^3 [1-p(t)]$. Since the update rule is  $(++-+) \rightarrow (-+-+)$, it contributes to the production of $\{+\}$ with a weight of $\frac{2}{4}=\frac{1}{2}$. Therefore its weighted contribution to the total update expression of $p(t+1$) is $\frac{1}{2} p(t)^3 [1-p(t)]$. Performing such an evaluation for each configuration of above sixteen ones and adding them,  we end up with Eq. (3)  where $m_{4,0}=1-m_{4,4}=0$, $m_{4,1}=1-m_{4,3}=\frac{1}{4} $ and $m_{4,2}=\frac{1}{2}$ which in turn yield exactly Eq. (5). Therefore it is  a Voter model as shown above. 

It could be argued that our scheme ignore the precise arrangement of the spins since we are using a probabilistic approach as opposed to a deterministic one. However that is not the case since we are considering the average result of deterministic rules to get the global symmetry  of repeated updates. As a proof of the validity of our transformation we can cite the  recent work by Behera and Schweitzer \cite{frank-voter} who first demonstrated the identity of  above Sznajd model and the Voter model using simulations and analytical calculations. We are recovering their result using our much simpler and general scheme. The analytical results of Mobilia and Redner for the case of size $3$ also confirms the validity of our simple approach \cite{red-2}.

On this basis we can shed a new light on original Sznajd finding that their simulations with an initial random distribution of $\{+\}$ and $\{-\}$ lead to a polarization along $\{+\}$ for $25\%$ of them, to to a polarization along $\{-\}$ for another $25\%$ and an anti-ferromagnetic like configuration with the last $50\%$ of configurations \cite{sznajd}. On the other hand, they noted that more initial $\{+\}$ leads to more polarization along $\{+\}$ than along $\{-\}$ and vice versa.  Behera and Schweitzer confirms these proportions from their simulations \cite{frank-voter}. 

From our finding, Sznajd rules [a, b, c, d] define a Voter model. In particular  the value of the parameter $m_{4,3}=\frac{3}{4}=0.75$. Accordingly an exact initial proportion of equal number of $\{+\}$ and $\{-\}$ should be preserved. However fluctuations in respectively the numbers of initial sites of each state,  their spatial distribution and the actual update process of selecting the pairs will lead to some very small departure of $m_{4,3}$ from $\frac{3}{4}=0.75$. On average statistics  produces half $m_{4,3}=\frac{3}{4}+\epsilon$ and half $m_{4,3}=\frac{3}{4}-\epsilon$. The $+\epsilon$ puts the dynamics in the polarization phase with half of the polarization along $\{+\}$ and the other half along  $\{-\}$ giving the two $25\%$. Naturally more initial $\{+\}$ leads to more polarizations along $\{+\}$. In contrast  the $-\epsilon$ drives the dynamics  into the consensus phase which has only one attractor   with an equal number of $\{+\}$ and $\{-\}$ resulting in the $50\%$ of anti-ferromagnetic like  runs.

Another version of Sznadj rules maintains the  rules for identical opinion at central pairs ([a] and [b] ) but in case of different opinions, it leaves  the two neighboring states unchanged instead of going anti-ferromagnetic like. It means to substitute [c]  and [d] with, \\

$[c']: (++-+),\ (++--),\ (-+-+),\ (-+--)\rightarrow $ the same,\\

$[d']: (+-++),\ (+-+-),\ (--++),\ (--+-)\rightarrow  $ the same.\\ \\
At this stage it is worth to notice that both versions are used indifferently in the literature as slightly different \cite{stauffer}. However performing the calculations of $p(t+1$) as above, we find $m_{4,0}=1-m_{4,4}=0$, $m_{4,1}=1-m_{4,3}=\frac{1}{8}$ and $m_{4,2}=\frac{1}{2}$ leading to, 
\begin{equation}
p(t+1)=p(t)^4+ \frac{7}{2} p(t)^3[1-p(t)]+3p(t)^2[1-p(t)]^2+\frac{1}{2}p(t)[1-p(t)]^3 \ ,
\end{equation}
which is different from Eq. (5). Here we have Eq. (4) with $m_{4,3}= \frac{7}{8}$ locating the case in the polarization phase since $\frac{7}{8}> \frac{3}{4}=\frac{6}{8}$. It implies polarization along the initial majority which again explains the Sznajd simulation finding that in this case they found only $\{+\}$ or $\{-\}$ ordering as expected from our framework.

A third version called  ``If you do not know what to do, just do nothing'' was briefly mentioned in the conclusion of the original Sznajd model  \cite{sznajd}. There, an agent takes the opinion of its two neighbors when they hold the same one. Otherwise it preserves it current state. The rules write $\{+-+\}\rightarrow\{+++\}$ and $\{-+-\}\rightarrow\{---\}$ with all other configurations being unchanged, i.e.,  $\{+--\}\rightarrow\{+--\}$, $\{--+\}\rightarrow\{--+\}$, $\{++-\}\rightarrow\{++-\}$, $\{-++\}\rightarrow\{-++\}$, $\{+++\}\rightarrow\{+++\}$, $\{---\}\rightarrow\{---\}$.

They noticed an enormous number of steady states. Within our scheme, it corresponds to groups of size $3$, i.e.,  $a_3=1$ with $m_{3,2}=\frac{7}{9}$. In this case, as found by Mobilia and Redner \cite{red-2}, the transition from polarization to consensus occurs at $m_{3,2}=\frac{2}{3}=\frac{6}{9}$ which thus puts the current case in the polarization phase with two attractors. However as seen in Fig. (2), the flow curve is very close to the Voter line ($\Delta m_{3,2} =\frac{1}{9}$) making the flow dynamics very slow. Accordingly their simulations were  either not long enough or of a too small size. 

We mention also a recent preprint by Sanchez \cite{sanchez} who suggests another modification of  Sznajd rules [c] and [d] by making each agent from the central pair to adopt its external neighbor state
 in order to avoid the $50\%$ cases of equal number of $\{+\}$ and $\{-\}$. Our scheme gives  $m_{4,0}=1-m_{4,4}=m_{4,1}=1-m_{4,3}=0$ with $m_{4,2}=\frac{1}{2}$. As seen from Fig (1) it  is exactly the Original Galam four size model with no bias for which indeed we are in the polarization phase for which indeed there exist only polarization along the initial majority.

\begin{figure}[h]
\begin{center}
\includegraphics[scale=1]{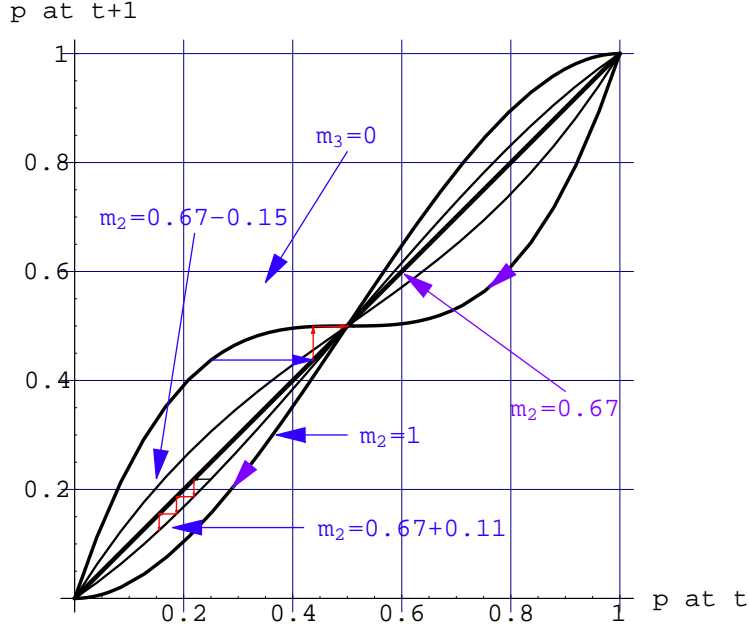}
\caption{Variation of the flow dynamics as function of $m_3^3$  with $a_3=1$ treated by Mobilia and Redner \cite{red-2}.  The transition occurs at $m_{3,2}=\frac{2}{3}=0.67$ where the Voter model is recovered. The cases $m_{3,2}=1$ (Original Galam) and  the extreme case of a systematic minority convincing power at $m_3^3=0$ are shown together with two intermediate cases below and above the critical case  at $m_{3,2}=\frac{2}{3}+\frac{1}{9 }$ and $m_{3,2}=\frac{2}{3}-0.15$. the $+\frac{1}{9 }=0.11$ corresponds to the third version of Sznajd model.}
\end{center}
\end{figure}

In a recent paper Slanina and H. Lavicka \cite{slanina} suggest to simplify Sznjad model in a similar way to above  last Sznjad version dealing with only three agents at a time. However instead of choosing the two external agents influencing the center one, they consider neighboring pairs to influence the third agent with $\{++-\}$ and $\{-++\}$ $\rightarrow\{+++\}$,  $\{--+\}$ and $\{+--\}$ $\rightarrow\{---\}$ . Other configurations stay unchanged with $\{+-+\}\rightarrow\{+-+\}$,   $\{-+-\}\rightarrow\{-+-\}$, $\{+++\}\rightarrow\{+++\}$, $\{---\}\rightarrow\{---\}$. Applying our scheme yields the value $m_{3,2}=\frac{8}{9}$. Contrary to the Sznajd third version which has $\Delta m_{3,2} =\frac{1}{9}$ from the transition Voter line, the Slanina and  Lavicka version has also a $\Delta m_{3,2} =\frac{1}{9}$ but from the original Galam voting model  \cite{voting-all}. Such a proximity shed a light on Slanina and Lavicka comment  ``It is rather interesting to observe that the deterministic dynamics of Galam model leads to a formula very similar to...'' \cite{slanina}.

In their paper Slanina and Lavicka \cite{slanina} also consider a simplified version of  Ochrombel model \cite{och} with only two agents at a time. The rules are $\{++\}\rightarrow\{++\}$, $\{--\}\rightarrow\{--\}$, $\{+-\}\rightarrow\{++\}$, $\{-+\}\rightarrow\{--\}$. They notice that it is a Voter model. Using our scheme 
we get $m_{2,2}=1=1-m_{2,0}=1$ and $m_{2,1}=\frac{1}{2}$ leading $p(t+1)=p(t)^2+p(t)[1-p(t)]=p(t)$ in agreement with their statement. It interesting to note that the original  Ochrombel model \cite{och} deals with 3 agents with the middle one influencing its two neighbors. Our scheme yields $p(t+1)=p(t)^3+2p(t)^2[1-p(t)]+p(t)[1-p(t)]^2=p(t)$ making $m_{3,1}=\frac{2}{3}$ which indeed locates the model at the critical line for groups of size $3$. Therefore, it is also a Voter model.

Last we can suggest another version of local rules which could be thought as opposing Sznajd social claim by making the core of influence to come from the border people instead of the center.  It is the border people who  convince the center ones if they are in the same state (inward) instead of going from center to border (outward ).  If they are in different states, nothing happens. Our scheme gives exactly the same results as Sznajd with rules [a, b, c', d'] with $m_{4,0}=1-m_{4,4}=0$, $m_{4,1}=1-m_{4,3}=\frac{1}{8}$ and $m_{4,2}=\frac{1}{2}$ leading to Eq. (6). Accordingly we conclude that inward and outward flow produces the same outcome as noticed earlier by Behera and Schweitzer \cite{frank-voter}.

To conclude, we have presented a new model which is shown to include most of two state local dynamics rules. It allows to define some symmetry to determine which states are true attractors of the corresponding dynamics. It could serve as guidelines to avoid possible misleading social and political claims from  incomplete simulations. Our reformulation can be extended to the Sznajd model at two dimension which considers groups of $8$ agents. Also asymmetric situations should be investigated. 

\section*{Acknowledgment}

I would like to thank D. Stauffer for nasty tough pertinent comments

%%%%%%%%%%%%%%%%%%%%%%


\begin{thebibliography}{88}

\bibitem {mino} S. Galam, Eur. Phys. J. B {\bf25} Rapid Note, 403 (2002)

\bibitem{contra} S. Galam, Physica  {\bf 333}, 453 (2004).

\bibitem {voting-all} S. Galam,  J. of Math. Psychology {\bf30}, 426  (1986);   J. of Stat. Phys. {\bf 61}, 943  (1990);   Physica A {\bf 285}, 66 (2000).

\bibitem {jan} S. Galam and J. P. Radomski, Phys. Rev. E { \bf 63}, 51907  (2001) 

\bibitem{red-1} P. L. Krapivsky and S. Redner, Phys. Rev. Lett. {\bf 90}, 238701 (2003).

\bibitem{red-2} M. Mobilia, S. Redner, Phys. Rev. E {\bf 68}, 046106  (2003).


\bibitem {socio}  S. Galam, Physica A {\bf 336}, 49 (2004)

\bibitem {strike}  S. Galam, Y. Gefen and Y. Shapir, Math. J. of Sociology  {\bf 9}, 1 (1982).   

\bibitem{frank-book} Modeling Complexity in Economic and Social Systems, ed. F. Schweitzer, World Scientific, Singapore  (2002).
 
 \bibitem {weidlich-book} W. Weidlich, Sociodynamics; A Systematic Approach to Mathematical Modelling in the Social Sciences. Harwood Academic Publishers  (2000).
 
 \bibitem{stauffer} D. Stauffer, Journal of Artificial Societies and Social Simulation {\bf 5} No. 1, paper 4 (2000) and AIP Conference Proceedings {\bf 690}147 (2003).
 
 \bibitem{neigbhor} C.J. Tessone, R. Toral, P. Amengual, H.S. Wio, and M. San Miguel
Eur. Phys. J. B {\bf 39},  535 (2004).

\bibitem{huber} F. Wu and B. A. Huberman,  arXiv: cond-mat/0407252

\bibitem{voter} T. M. Liggett, Stochastic Interacting Systems: Contact, Voter, and Exclusion Processes, (Springer, Berlin (1999).

\bibitem{sznajd} K. Sznajd-Weron and J. Sznajd, Int. J. Mod. Phys. C {\bf 11}, 1157 (2000).

\bibitem{frank-voter} L. Behera and F. Schweitzer,  Int. J. Mod. Phys. C {\bf 14}, 1331(2003)

\bibitem{mosco-1}  S. Galam and S. Moscovici, Euro. J. of Social Psy. {\bf 21}, 49 (1991).  

\bibitem{sanchez} J.  R. S«anchez, arXiv: cond-mat/0408518 

\bibitem{slanina} F. Slanina and H. Lavicka, Eur. Phys. J. B {\bf  35}, 279 (2003).

\bibitem{och} R. Ochrombel, Int. J. Mod. Phys. C {\bf 12}, 1091 (2001).

\end{thebibliography}
 \end{document}